\documentclass[runningheads]{llncs}

\usepackage[T1]{fontenc}
\usepackage{graphicx}
\usepackage{amsmath}
\usepackage{booktabs}
\usepackage{multirow}
\usepackage{xspace}
\usepackage{url}
\usepackage{tikz}
\usepackage{bbding}
\usetikzlibrary{arrows.meta,positioning,fit,backgrounds}
\setkeys{Gin}{draft=false}

\newcommand{\system}{\textsc{FastKVComp}\xspace}
\newcommand{\kvserve}{\textsc{KVServe}\xspace}
\newcommand{\tilelang}{\textsc{TileLang}\xspace}
\newcommand{\lc}{\textsc{LC}\xspace}

\begin{document}

\title{Beyond Compression Ratio: High-Throughput KV Compression for Disaggregated LLM Serving}

\titlerunning{High-Throughput KV Cache Compression}

\newcommand{\equalcontrib}{\textsuperscript{*}}

\author{
Junhong Wu\inst{1,2}\equalcontrib \and
Zedong Liu\inst{1,2}\equalcontrib \and
Xinyang Ma\inst{1,2} \and
Yuhan Chen\inst{1,2} \and \\
Guangming Tan\inst{1} \and
Dingwen Tao\inst{2}\Envelope
}

\authorrunning{J. Wu and Z. Liu et al.}

\institute{
Institute of Computing Technology, Chinese Academy of Sciences
\\
\and
University of Chinese Academy of Sciences\\
\email{taodingwen@ict.ac.cn}
}

\maketitle

\begingroup
\renewcommand{\thefootnote}{*}
\footnotetext{These authors contributed equally.}
\endgroup

\begin{abstract}
KV cache compression reduces the communication volume of disaggregated
large language model serving, but its practical benefit depends not only
on compression ratio and model quality, but also on how efficiently the
compression pipeline executes. This requirement becomes increasingly
important as modern serving systems adopt high-bandwidth interconnects,
where inefficient compression can offset a substantial portion of the
saved communication time.

This paper presents \system, a high-throughput KV cache compression
data path built on top of \kvserve. \system improves compression
execution in two complementary ways. First, it fuses the distribution
transformation and low-bit quantization stages into a KV-aware
\tilelang kernel, eliminating intermediate tensor materialization and
redundant memory accesses. Second, it replaces the
original lossless compression backend with an open, extensible, and
searchable backend based on \lc, enabling high-performance compression
operator pipelines. The fused kernel achieves up
to $5.9\times$ speedup, while the new lossless backend increases
compression throughput from below 90~GB/s to over 500~GB/s. Across the evaluated
workloads, \system reduces job completion time by up to $2.04\times$.
End-to-end evaluation shows that these optimizations allow KV
compression to retain meaningful communication benefits under modern
high-bandwidth network.
\keywords{KV cache compression \and disaggregated LLM serving \and
operator fusion \and lossless compression \and GPU systems}
\end{abstract}

\section{Introduction}
\label{sec:introduction}

Disaggregated large language model (LLM) serving increasingly separates
computation and storage responsibilities across devices and nodes.
Representative scenarios include prefill--decode disaggregation, remote
KV cache offloading~\cite{splitwise2024,mooncake2025,cachegen2024}. In these systems,
the key-value (KV) cache is no longer only local inference state; it also
becomes a major communication object whose volume grows with model size,
batch size, and context length.

Recent KV cache compression methods reduce this communication volume
through low-bit quantization, mixed-precision representations,
distribution transformations, and lossless encoding~\cite{kivi2024,kvquant2024,cachegen2024,kvserve2026}. Most existing
studies focus primarily on the tradeoff among compression ratio, model
quality, and workload-specific compression policies. These studies show
that KV caches are highly compressible, but they do not fully address
the efficiency of executing the resulting compression pipeline.

Compression reduces end-to-end communication latency only when the
saved transfer time is larger than the encoding and decoding overhead.
This condition becomes more demanding as serving systems adopt
100--400~Gbps networks and high-performance remote direct memory access
(RDMA) data paths~\cite{splitwise2024,mooncake2025,kvserve2026}. Reducing KV volume is therefore not sufficient:
compression must also execute fast enough to keep pace with modern
serving networks.

We identify two major optimization opportunities in the execution path
of practical KV cache compression. First, distribution transformation
and quantization are commonly implemented as separate GPU kernels.
Their intermediate tensor is written to high-bandwidth memory (HBM) and
immediately read by the following quantization kernel, introducing
redundant memory traffic, temporary storage, and kernel-launch overhead.
Second, the original lossless compression backend used by
\kvserve~\cite{kvserve2026} provides limited openness and extensibility
for integrating and combining high-performance compression primitives. Its effective throughput is therefore insufficient to fully exploit high-bandwidth
serving data paths.

To address these issues, we present \system, a high-throughput KV cache
compression data path integrated into \kvserve. \system fuses
transformation and quantization into a single KV-aware \tilelang kernel~\cite{tilelang2025},
avoiding the materialization of intermediate transformed tensors. It
also introduces an open and searchable lossless compression backend
based on \lc~\cite{lcframework2025,azami2025lossless}, which composes high-performance primitives into efficient
compression pipelines for quantized KV data. These optimizations
preserve the compression policies and model-quality characteristics of
\kvserve while substantially improving their execution efficiency.

Our preliminary implementation achieves up to $5.9\times$ acceleration
for fused transformation and quantization. The new lossless compression
backend improves effective throughput from below 90~GB/s to
over 500~GB/s.

In summary, this paper makes the following contributions:

\begin{itemize}
    \item We characterize the execution bottlenecks of practical KV cache
    compression, revealing substantial overhead from fragmented GPU
    operators and the inflexible lossless compression path.

    \item We design a KV-aware \tilelang frontend that fuses distribution
    transformation and low-bit quantization, while optimizing the inverse
    reconstruction path through tiled execution and workspace reuse.

    \item We develop an open and searchable lossless compression backend
    based on \lc, which composes GPU primitives into pipelines optimized
    for end-to-end KV transfer.

    \item We integrate the optimized data path into \kvserve and evaluate
    it on both host-staged and cross-node InfiniBand deployments, achieving
    up to $2.04\times$ JCT speedup and $2.01\times$ throughput improvement
    over the original \kvserve implementation.
\end{itemize}

\section{Background and Gap Analysis}
\label{sec:background}

\subsection{KV Transfer and Compression Pipeline}
\label{sec:compression-pipeline}

\textbf{KV Cache Compression.}
KV caches grow linearly with context length and can occupy substantial
memory for long-context requests. Existing compression methods reduce
this footprint through low-bit quantization, mixed-precision
representations, distribution transformations, and lossless
coding~\cite{cachegen2024}. While originally motivated largely by
memory capacity, these techniques are increasingly important when KV
states must move across devices or nodes.

\textbf{Compression in KV Transfer.}
In disaggregated serving, KV caches are transferred between prefill and
decode workers, offloaded to remote memory or storage, and reused across
requests or serving sessions~\cite{mooncake2025,cachegen2024}.
These scenarios turn the KV cache into a large communication payload,
often reaching gigabytes for large models and long contexts. Systems
such as \kvserve~\cite{kvserve2026} therefore apply compression directly to the KV transfer
path, reducing communication volume while preserving serving quality.

A typical compression path first transforms the KV distribution,
quantizes the transformed values into a compact low-bit representation,
and optionally applies lossless compression before transfer. The
receiver performs the inverse operations to reconstruct the KV states
required by attention. For an original payload of size
$D_{\mathrm{raw}}$, compressed size $D_{\mathrm{comp}}$, and effective
network bandwidth $B_{\mathrm{net}}$, compression is beneficial when
the saved transfer time exceeds its encoding and decoding overhead:
\begin{equation}
\small
T_{\mathrm{encode}} + T_{\mathrm{decode}}
<
\frac{D_{\mathrm{raw}}-D_{\mathrm{comp}}}{B_{\mathrm{net}}}.
\label{eq:benefit-condition}
\end{equation}

As network bandwidth increases, this condition leaves progressively
less room for compression overhead. Practical KV compression must
therefore optimize not only compressed size, but also the efficiency of
the compression data path itself.

\subsection{Optimization Opportunities}
\label{sec:opportunities}

\subsubsection{Fragmented Transformation and Quantization.}
In the original execution path, transformation and quantization are
implemented as separate GPU kernels. The transformation kernel reads
the original KV tensor and writes a full intermediate tensor to HBM.
The quantization kernel then reads this intermediate tensor, computes
or applies the quantization parameters, and writes the low-bit output.

This design introduces two avoidable costs. First, the intermediate
tensor incurs an additional HBM write and read, even though it is
consumed only by the immediately following quantization stage. Second,
separate kernels introduce additional launch and synchronization
overheads. These costs are especially important because both stages are
closely coupled and process the same KV tensor layout.

\subsubsection{Limited Lossless Compression Extensibility.}
The original \kvserve implementation uses nvCOMP through its available
integration path. Although nvCOMP provides several GPU compression
algorithms, the original integration offers limited flexibility for
inspecting, extending, and composing the underlying compression
operations. In particular, it is difficult to incorporate new
high-performance primitives or construct specialized operator
pipelines for quantized KV data.

As a result, the effective end-to-end throughput of the original
lossless compression stage remains below 100~GB/s in our initial
measurements. This leaves substantial room for improving the execution
efficiency of KV compression, particularly for systems equipped with
modern high-bandwidth interconnects.

Based on these observations, we establish two design goals:

\begin{itemize}
    \item \textbf{G1: Eliminate redundant memory traffic.}
    Closely coupled transformation and quantization operations should
    execute within a single GPU kernel whenever possible.

    \item \textbf{G2: Build a high-throughput and extensible backend.}
    The lossless compression layer should expose composable
    high-performance primitives and support efficient pipeline
    construction for quantized KV data.
\end{itemize}

\section{Design Overview}
\label{sec:overview}

\begin{figure}[!t]
    \centering
    \includegraphics[
        draft=false,
        width=0.94\linewidth,
        keepaspectratio
    ]{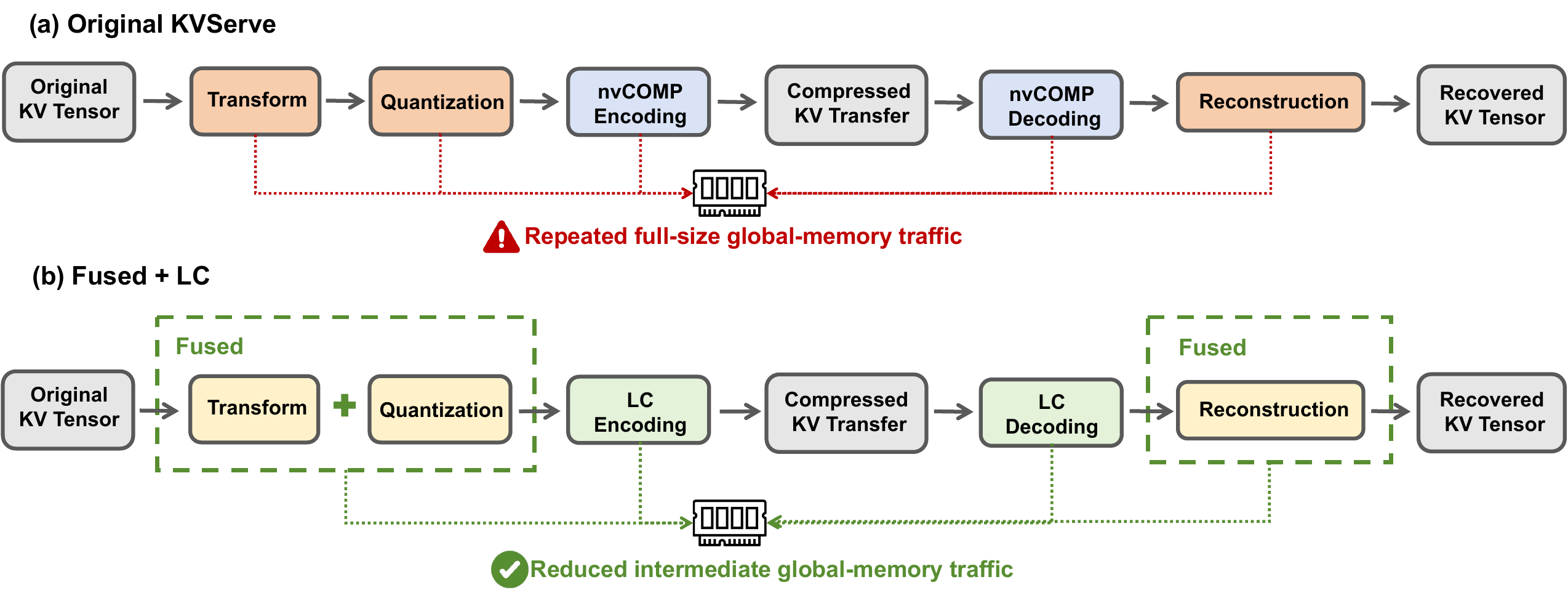}
    \caption{Original KVServe and the proposed \system execution paths.}
    \label{fig:system-overview}
\end{figure}

\system is a high-throughput implementation of the KV compression data
path below the compression-policy layer of KVServe. It preserves the
original strategy-selection and compression semantics while optimizing
how a selected strategy is executed.

\system consists of two components: a tiled \tilelang frontend that
streamlines transformation, quantization, and reconstruction using
reusable GPU workspaces, and an \lc-based lossless backend that provides
composable compression primitives and offline pipeline search. At the
sender, quantized KV data is written directly into a preallocated output
buffer before lossless encoding. At the receiver, lossless decoding is
followed by the optimized reconstruction path. The design prioritizes
high throughput, low execution overhead, compatibility with existing
\kvserve policies, and extensibility to new compression components.

\section{Ultra-fast KV Compression}
\label{sec:design}

\subsection{Fused Transform--Quantization (Addressing G1)}
\label{sec:fused-kernel}

To eliminate the redundant memory traffic identified as G1, we fuse
transformation and quantization into a single tiled \tilelang kernel
(D1). The fused frontend combines the randomized Hadamard transformation,
quantization-statistics computation, and low-bit conversion within a
single GPU kernel.

Each KV tile is loaded once and processed using on-chip storage.
Intermediate transformed values are directly consumed by the
quantization stage instead of being materialized as a full-size tensor
in global memory. The operator then writes the quantized representation
directly into a preallocated all-layer output buffer, reducing temporary
allocation, head splitting, per-layer stacking, layout conversion, and
framework-level kernel-launch overhead.

The fused operator preserves the original \kvserve compression policy,
including the deterministic randomized Hadamard transformation,
head-importance-aware mixed precision, channel-wise quantization for
keys, token-wise quantization for values, and the original min--max
quantization rule. Transformation signs, precision assignments,
compiled \tilelang operators, and auxiliary workspaces are cached and
reused across layers and requests.

\subsubsection{Fused Reconstruction.}

On the receiver, the TileLang reconstruction operator combines
dequantization, inverse Hadamard transformation, normalization, and sign
restoration in a tiled execution pipeline. It avoids materializing a
full-size dequantized intermediate tensor and writes only the
reconstructed KV representation to global memory. This optimization is
independent of the selected lossless backend and therefore applies to
both nvCOMP and LC.

\subsection{Searchable High-Throughput Lossless Backend (Addressing G2)}
\label{sec:lossless-backend}

To meet G2, we replace the fixed nvCOMP path with a searchable
high-throughput lossless backend based on \lc (D2). nvCOMP provides a readily
available GPU-resident codec and achieves a high compression ratio. Such a design
is effective when network transfer dominates the KV delivery path,
because the reduction in transferred data can compensate for the
additional encoding and decoding cost.

However, this tradeoff changes under high-bandwidth communication and
after frontend optimization. Once transformation and quantization
overhead is reduced, lossless encoding and decoding account for a
larger fraction of the remaining critical path. Moreover, nvCOMP is
distributed as a binary library with a fixed public interface, which
limits the ability to inspect, recombine, or specialize its internal
operations for low-bit KV data.

Unlike a fixed general-purpose codec, \lc~\cite{lcframework2025,azami2025lossless}
constructs compression algorithms from composable GPU primitives,
including tuple reordering, bit shuffling, and redundancy encoding.
This allows the backend to be specialized for the statistical
properties of quantized KV buffers. It also enables pipeline selection
to consider encoding latency, decoding latency, and compressed size
jointly rather than optimizing compression ratio alone.

\subsubsection{Offline Pipeline Search.}
\label{sec:lc-search}

We evaluate candidate two- and three-stage LC pipelines offline using
quantized KV buffers generated by the optimized frontend. For each
candidate, we measure GPU encoding time, GPU decoding time, and
compressed size. We then retain the Pareto-efficient candidates and
rank them for the evaluated high-bandwidth communication setting.

This procedure selects
\texttt{TUPL8\_1}$\rightarrow$\texttt{BIT\_8}$\rightarrow$
\texttt{RZE\_2}. Tuple reordering and bit shuffling expose regularities
in the low-bit KV representation, while zero-oriented encoding removes
the resulting redundancy. Although the selected LC pipeline provides a
lower compression ratio than nvCOMP ANS, its substantially lower
encoding and decoding latency makes it competitive on the end-to-end
critical path. Once selected, the pipeline remains fixed during online
serving, so no search is performed at runtime.

\subsubsection{Native GPU Integration.}

We generate an in-process GPU runtime specialized for the selected LC
pipeline. The runtime directly consumes the CUDA buffer produced by the
fused TileLang frontend and returns a GPU-resident compressed bitstream.
The online serving path therefore avoids LC's file-based command-line
interface, host-side payload serialization, and CPU--GPU movement of
the compressed payload. Only a small compressed-size scalar is returned
to the host during encoding.

The runtime exposes device-pointer encoding and decoding APIs directly
to the KV transfer path. Persistent runtime scratch space and a reusable
encode staging buffer reduce repeated allocation on the critical path.
The component-level effects of this integration are evaluated in
Sec.~\ref{sec:component-results}.

\subsubsection{Extensibility.}

Because LC is component based, additional transformations, encoders,
and architecture-specific implementations can be incorporated into the
offline search without modifying the upper-level \kvserve control
logic. The same selection method can therefore adapt the lossless
backend to future KV formats, models, communication environments, and
GPU architectures.

D1 removes the dominant transform/quantization overhead (G1), while D2
prevents lossless coding from becoming the next bottleneck (G2); together
they keep the whole path fast enough for high-bandwidth transfer.

\begin{figure}[!t]
    \centering
    \includegraphics[
        draft=false,
        width=0.90\linewidth,
        keepaspectratio
    ]{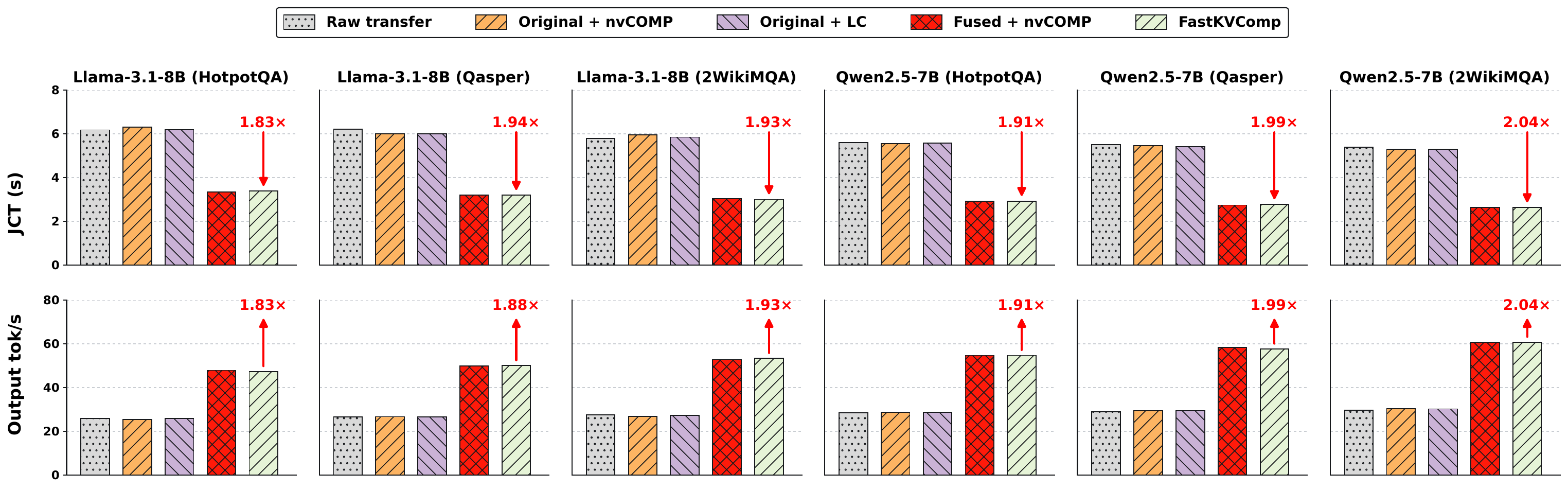}
    \caption{End-to-end results over the host-staged NCCL SHM path.}
    \label{fig:e2e-shm-results}
\end{figure}

\section{Evaluation}
\label{sec:evaluation}

We evaluate \system in terms of end-to-end serving performance,
component-level efficiency, and the contribution of individual
optimizations.

\paragraph{Experimental setup.}
Our local experiments use two NVIDIA RTX~5090 GPUs (32\,GB each) for
prefill and decode, respectively, without tensor parallelism. KV
payloads are transferred through the host-staged NCCL SHM path over
PCIe Gen5 x16, with measured bandwidth up to 306.5~Gbps. The software
stack includes CUDA~12.8, PyTorch~2.10.0, and
vLLM~0.18.0~\cite{vllm2023}.

We additionally evaluate a two-node deployment connected by 200-Gb/s
InfiniBand, using an NVIDIA RTX PRO~6000 for prefill and an NVIDIA H100
for decode. KV caches are compressed on the prefill node, transferred
over InfiniBand, and reconstructed on the decode node.

\paragraph{Workloads.}
We evaluate Llama-3.1-8B-Instruct~\cite{llama3report} and
Qwen2.5-7B-Instruct~\cite{qwen25report} on HotpotQA, Qasper, and
2WikiMQA from LongBench~\cite{longbench2024}. Inputs contain up to
20K tokens under a 32K context window. Each configuration processes
50 requests, generating 8 tokens locally and up to 64 tokens in the
remote deployment. Compilation, initialization, and warm-up costs are
excluded.

\paragraph{Baselines and metrics.}
We compare \textbf{Raw Transfer}, \textbf{Original KVServe},
\textbf{Fused Only}, \textbf{LC Only}, and \textbf{\system}.
All compressed configurations use identical compression policies,
isolating the effects of operator fusion and the lossless backend.
We report JCT, output throughput, KV transfer cost, compression ratio,
and component-level compression and reconstruction overhead.

\subsection{End-to-End Results}
\label{sec:end-to-end}

Figure~\ref{fig:e2e-shm-results} reports the end-to-end performance over
the host-staged NCCL SHM path.
Compared with Raw Transfer, \system improves output throughput by
1.83--2.04$\times$ and reduces JCT by 45.3--51.1\% across both models
and all workloads. Over Original KVServe, throughput improves by
1.87--2.01$\times$. Although Original KVServe reduces KV volume by
9.4--10.1$\times$, its fragmented compression path offsets most of the
communication savings. In contrast, the fused configurations nearly
double throughput, showing that efficient compression execution is
critical at high network bandwidths.

\begin{figure}[!t]
    \centering
    \includegraphics[
        draft=false,
        width=0.94\linewidth,
        keepaspectratio
    ]{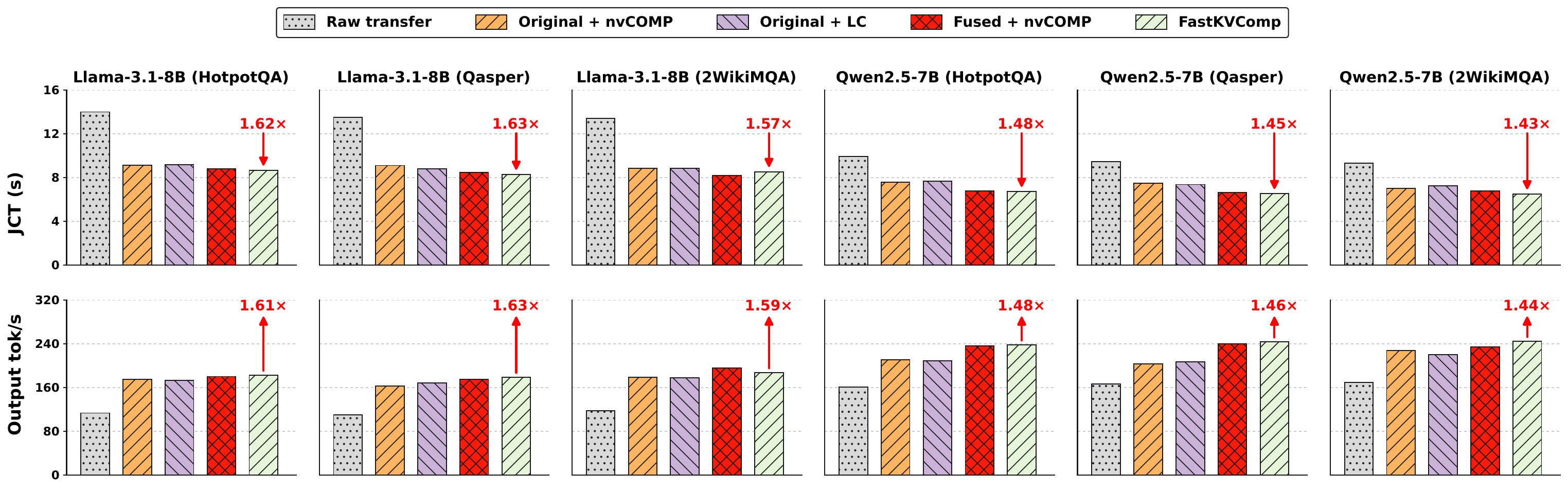}
    \caption{End-to-end JCT and output throughput over 200-Gb/s
    InfiniBand. Red annotations show the improvement of \system over
    Raw Transfer.}
    \label{fig:e2e-ib-results}
\end{figure}

Figure~\ref{fig:e2e-ib-results} further evaluates \system in the two-node 200-Gb/s InfiniBand deployment. Across the remote workloads, \system reduces JCT by 30.2--38.6\% and
improves output throughput by 44.2--62.6\% over Raw Transfer. These
results confirm that the optimized compression path remains effective
in heterogeneous, high-bandwidth cross-node serving.
\subsection{Component-Level Results}
\label{sec:component-results}

We isolate the costs of the fused frontend and the lossless backend
using two representative KV requests.

\subsubsection{Fused Transform--Quantization.}

The original transformation--quantization path takes
27.309--34.394~ms, whereas the fused TileLang implementation reduces
this to 5.876--6.926~ms, yielding a 3.94--5.85$\times$ speedup and
74.6--82.9\% lower preprocessing latency. The gain comes primarily from
eliminating full-size intermediate KV materialization and reusing tiled
workspaces, while preserving the original transformation and
quantization semantics.

\subsubsection{Lossless Compression Backend.}

Using the LC pipeline selected in Sec.~\ref{sec:lc-search}, lossless
encoding latency drops from 3.179--3.274~ms with nvCOMP ANS to
0.443--0.531~ms, corresponding to a 6.0--7.4$\times$ speedup. Decoding
similarly decreases from 2.543--2.666~ms to 0.488--0.575~ms, or about
4.9$\times$ on average.

\begin{figure}[!t]
    \centering
    \includegraphics[
        draft=false,
        width=0.86\linewidth,
        height=0.52\textheight,
        keepaspectratio
    ]{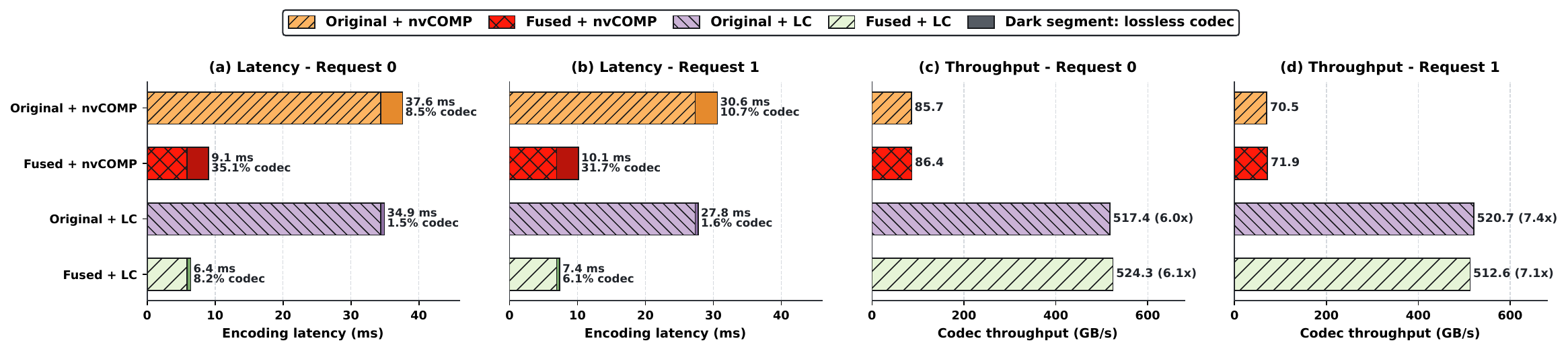}
    \caption{Encoding-time breakdown for the original and fused
    frontends using nvCOMP and the searched LC pipeline.}
    \label{fig:lc-bottleneck-breakdown}
\end{figure}

As shown in Fig.~\ref{fig:lc-bottleneck-breakdown}, after frontend
fusion, lossless encoding accounts for 31.7--35.1\% of the encoding
path with nvCOMP but only 6.1--8.2\% with LC. This lower codec overhead
comes at a moderate compression-ratio tradeoff: 8.850--9.168$\times$
for nvCOMP versus 6.344--6.360$\times$ for LC.

\subsubsection{Complete Compression Path.}

Table~\ref{tab:hardware-compression-summary} summarizes the complete
compression and reconstruction paths on the GPUs used in the remote
deployment. The critical transfer direction compresses KV caches on
the RTX PRO~6000 prefill node and reconstructs them on the H100 decode
node; results on both GPUs are reported for completeness.

\begin{table}[!t]
    \caption{Component-level compression and decompression performance
    in the 200-Gb/s InfiniBand deployment. The RTX PRO 6000 and H100
    serve as the prefill and decode GPUs, respectively.}
    \label{tab:hardware-compression-summary}
    \centering
    \scriptsize
    \setlength{\tabcolsep}{3pt}
    \renewcommand{\arraystretch}{1.12}
    \resizebox{\linewidth}{!}{%
    \begin{tabular}{llccccc}
        \toprule
        GPU &
        Path &
        KV comp. ratio &
        Comp. throughput &
        Comp. latency &
        Decomp. throughput &
        Decomp. latency \\
        & & $(\times)$ & (GB/s) & (ms) & (GB/s) & (ms) \\
        \midrule

        RTX PRO 6000 &
        Original + nvCOMP &
        \textbf{8.867} &
        34.773 &
        30.879 &
        32.284 &
        33.260 \\

        RTX PRO 6000 &
        Original + LC &
        6.474 &
        41.591 &
        25.817 &
        37.336 &
        28.759 \\

        RTX PRO 6000 &
        Fused + LC &
        6.276 &
        \textbf{139.218} &
        \textbf{7.713} &
        \textbf{155.841} &
        \textbf{6.890} \\

        \midrule

        H100 &
        Original + nvCOMP &
        \textbf{8.804} &
        27.659 &
        38.820 &
        26.698 &
        40.219 \\

        H100 &
        Original + LC &
        6.478 &
        30.195 &
        35.560 &
        29.038 &
        36.977 \\

        H100 &
        Fused + LC &
        6.280 &
        \textbf{91.606} &
        \textbf{11.721} &
        \textbf{103.104} &
        \textbf{10.414} \\

        \bottomrule
    \end{tabular}%
    }
\end{table}

On the RTX PRO~6000 prefill GPU, \system reaches
139.218~GB/s compression throughput with 7.713~ms latency, providing a
4.00$\times$ throughput improvement and 75.0\% latency reduction over
Original KVServe. On the H100 decode GPU, it reaches 103.104~GB/s
decompression throughput with 10.414~ms latency, corresponding to a
3.86$\times$ throughput improvement and 74.1\% lower latency. These
results demonstrate substantial overhead reduction on both ends of the
remote KV-transfer path.

For the two representative requests, the complete encoding path drops
from 30.583--37.599~ms with Original KVServe to 6.400--7.376~ms with
\system, yielding a 4.15--5.88$\times$ speedup and 75.9--83.0\%
latency reduction. Together, the component-level results explain the
end-to-end gains: operator fusion removes the dominant frontend
overhead, while LC prevents lossless coding from becoming the next
bottleneck.

\subsection{Ablation Study}
\label{sec:ablation}

Across both models and all three datasets, Fused Only improves output
throughput by 1.87--2.01$\times$ over Original KVServe, confirming
operator fusion as the primary source of end-to-end improvement under
the evaluated SHM bandwidth. LC Only remains within 2.1\% of Original
KVServe: its lower codec latency is largely offset by the increased
transfer volume from a lower compression ratio.

Fused Only and \system differ by at most 1.3\% in output throughput.
The former benefits from the higher compression ratio of nvCOMP,
whereas \system favors the substantially lower codec overhead of LC.
This tradeoff shows that the optimal lossless backend depends on the
target communication regime rather than compression ratio alone.

\section{Related Work}
\label{sec:related-work}

\subsubsection{KV Cache Compression}
\label{sec:related-kv}
Existing KV cache compression methods reduce memory and communication
costs through low-bit quantization~\cite{kivi2024,kvquant2024,gear2024},
mixed precision~\cite{mikv2024}, distribution
transformations~\cite{spinquant2025}, sparsification~\cite{h2o2023,snapkv2024},
and lossless coding~\cite{cachegen2024,kvserve2026}. They improve the
compression--quality tradeoff by reshaping KV distributions, allocating
precision, or exploiting redundancy in quantized
representations~\cite{kvbenchmark2024}. Recent systems combine KV compression with
runtime management or GPU-friendly lossless coding~\cite{splitzip2026},
but complete pipeline efficiency remains less explored. \system is
complementary: it preserves the quantization policy selected by \kvserve
while optimizing its execution.

\subsubsection{Operator Fusion}
\label{sec:related-fusion}

Operator fusion reduces kernel launches, intermediate tensors, and
memory traffic. Frameworks such as Triton~\cite{triton2019}, \tilelang,
and TVM~\cite{tvm2018} support specialized fused kernels. \system applies
these principles to the tightly coupled transformation--quantization
pair, consuming transformed values within one tiled GPU kernel without
global-memory materialization.

\section{Conclusion}
\label{sec:conclusion}

Compression throughput is a first-class metric for KV cache compression
in disaggregated LLM serving. \system combines a KV-aware fused
transformation--quantization kernel with an open, searchable lossless
backend, addressing overhead in both the frontend and codec. The fused
kernel achieves up to $5.9\times$ speedup, while the new backend exceeds
500~GB/s.These optimizations preserve the end-to-end benefits of KV
compression over high-bandwidth networks.
\begin{credits}
\subsubsection{\ackname}
This work was supported by the National Natural Science Foundation of
China through the Excellent Young Scientists Fund Program (Overseas)
(Project No.~E401470), the National Key Research and Development Program
of China (Grant Nos.~2020YFA0907000 and 2025YFB30037002), the National
Natural Science Foundation of China (Grant Nos.~62032023 and T2125013),
and Ant Group. The AI-driven experiments, simulations, and model training
were performed on the robotic AI-Scientist platform of the Chinese
Academy of Sciences.

\subsubsection{\discintname}
The authors have no competing interests to declare that are relevant
to the content of this article.
\end{credits}

\bibliographystyle{splncs04}
\bibliography{references}

\end{document}